\def\ref{\par\noindent\hang}

\def\ApJ{{\em Astrophys.~J.}}

\def\MN{{\em Mon.~Not.~R.~astr.~Soc.}}
\def\Nat{{\em Nature}}
\def\PASJ{{\em Publ.~astr.~Soc.~Japan}}

\def\spose#1{\hbox to 0pt{#1\hss}}
\def\approxlt{\mathrel{\spose{\lower 3pt\hbox{$\sim$}}
	\raise 2.0pt\hbox{$<$}}}
\def\approxgt{\mathrel{\spose{\lower 3pt\hbox{$\sim$}}
	\raise 2.0pt\hbox{$>$}}}

\def\multleft#1{\hbox to size{\vbox {\halign {\lft{##}\cr #1}}\hfill}\par}
\def\multright#1{\hbox to size{\vbox {\halign {\rt{##}\cr #1}}\hfill}\par}

\def\today{\ifcase\month\or January\or February\or March\or April\or May\or
      June\or July\or August\or September\or October\or November\or December\fi
      \space\number\day, \number\year}
\def\$<${\thinspace}

\def\boxit#1{\vbox{\hrule\hbox{\vrule\kern3pt\vbox{\kern3pt
          #1 \kern3pt}\kern3pt\vrule}\hrule}}

\documentstyle[psfig]{mn}

\title[Iron line emission from within the innermost stable orbit]
{Iron line profiles including emission from within the innermost
stable orbit of a black hole accretion disc}
\author[Young et al.]
{\parbox[]{6.in}{A.J. Young$^1$, R.R. Ross$^{1,2}$ and A.C.
Fabian$^1$ \\
\footnotesize
$^1$\emph{Institute of Astronomy, Madingley Road, Cambridge CB3 0HA}\\
$^2$\emph{Department of Physics, College of the Holy Cross, Worcester, MA}}}

\begin{document}

\maketitle

\begin{abstract}
Reynolds \& Begelman (1997) have recently proposed a model in which
the broad and extremely redshifted iron line seen during a deep
minimum of the light curve of the Seyfert 1 galaxy MCG--6-30-15
originates from matter spiralling into a Schwarzschild black hole,
contrary to previous claims that the black hole may be spinning
rapidly (Iwasawa et al 1996; Dabrowski et al 1997). Here we calculate
in detail the X-ray spectrum produced by their model using the full
reflected continuum emission, including absorption features. This
calculation takes into account the doppler and relativistic
effects. For the range of parameters we consider, we find that the
spectrum should show a large photoelectric absorption edge of iron,
which is not seen in the data. The absorption edge is a consequence of
the line emitting matter within the innermost stable orbit being
highly ionized, and is largely independent of the parameters chosen
for their model. If we restrict our attention to the 3--10 keV band we
may effectively remove this absorption edge by fitting a steeper power
law, but this results in a significant underprediction of the 0.4--0.5
keV flux. We conclude that the data on MCG--6-30-15 are more
consistent with the Kerr than the Schwarzschild model.
\end{abstract}


\section{Introduction}
Iwasawa et al (1996) have found tentative evidence that the black hole
in the Seyfert 1 galaxy MCG--6-30-15 is spinning rapidly. The iron
K$\alpha$ emission line in that source (Tanaka et al 1995) was seen to
shift to lower energies during a deep minimum of the light curve, and
to have an extremely large equivalent width $\sim 1$ keV. The iron
line emission is thought to originate from fluorescence in the
accreting matter, and the photons reaching a distant observer are
doppler and transverse-doppler shifted due to the motion of the
matter, and gravitationally redshifted in escaping the deep potential
well of the black hole. In the case of an accretion disc around a
Schwarzschild black hole stable circular orbits can only exist beyond
the radius of marginal stability, $r_{ms}=6m=6GM/c^2$. The
observations require a greater redshift than can be explained by
fluorescence from matter in a disc extending in to this radius.  A
disc around a spinning Kerr black hole, however, can extend further in
to regions where the gravitational redshift is stronger. Such a disc
can provide the necessary additional redshifted flux. A study by
Dabrowski et al (1997) has shown that, if this is the case, then the
data require that the specific angular momentum of the black hole
exceeds 0.95.

Recently Reynolds \& Begelman (1997) have proposed an alternative
model, hereafter RB97, in which the necessary additional redshifted
flux is provided by fluorescence in optically-thick ionized matter
spiralling in from the innermost stable orbit around a Schwarzschild
black hole. If this can provide a satisfactory fit to the data then
any conclusions on black hole spin made from present X-ray iron line
profiles are ambiguous. As they note, a rapidly spinning black hole in
a radio-quiet AGN (such as that implied by the maximal Kerr result for
MCG--6-30-15) is contrary to some schemes for explaining radio
loudness (e.g. Rees et al 1982; Wilson \& Colbert 1995).

In the RB97 model the accretion disc is assumed to extend in to the
radius of marginal stability, $6m$, within which matter is freely
falling into the black hole. The matter free-falling within $6m$ does
not dissipate any gravitational potential energy, and is unable to
support its own corona. To produce fluorescence in this matter it
needs to be illuminated by some external X-ray source, which is
approximated by a point source along the rotation axis of the disc. As
the material falls from $6m$ its density drops rapidly and it becomes
highly ionized as it approaches the black hole. The fluorescent line
that the illuminated matter produces is very dependent upon its
ionization state and a useful quantity is the ionization parameter
$\xi(r)=4\pi F_{\rm x}(r)/n(r)$, where $F_{\rm x}(r)$ is the X-ray
flux received per unit area of the disc at a radius $r$, and $n(r)$ is
the comoving electron number density. The iron line emission for
various ionization parameters has been investigated by Matt et al
(1993, 1996) and they conclude the following for various values of
$\xi$.

\newcounter{fig}
\begin{list}{\arabic{fig}.}{\usecounter{fig}}
\item $\xi <100$ ergs cm s$^{-1}$ the material produces a `cold' iron
line at 6.4 keV and only a small absorption edge.
\item 100 ergs cm s$^{-1}<\xi <500$ ergs cm s$^{-1}$ does not produce
an iron line because photons near the line energy are resonantly
trapped and lost due to Auger ejections. There is a moderate
absorption edge.
\item 500 ergs cm s$^{-1}<\xi <5000$ ergs cm s$^{-1}$ produces a `hot'
iron line at 6.8 keV with twice the fluorescent yield of the cold
line. There is a large absorption edge.
\item $\xi >5000$ ergs cm s$^{-1}$ does not produce an iron line because
the iron is completely ionized. There is no absorption edge.
\end{list}

As the matter falls toward the black hole its ionization parameter
increases monotonically, so there will be four regions within $6m$,
each identified with one of the ranges of ionization parameter given
above. Fig.~1 shows the typical behaviour of $\xi$, for the RB97
model, within $6m$. The first region just within $6m$ is very small
since the ionization parameter quickly exceeds $100$ ergs cm s$^{-1}$,
and there will be an annulus of material further in that may produce a
hot line. It is the emission from this region that provides the highly
redshifted flux that is required to fit the data. The reflected
continuum produced in this region necessarily has an absorption edge
associated with it, and it is important to consider the effect of this
on the predicted spectrum. In computing the line profile they expect
from their model Reynolds \& Begelman (1997) treat the iron emission
line as a delta function in the rest frame of the gas, which is then
broadened by doppler and relativistic effects. Here we use the entire
reflected X-ray spectrum, including the absorption edge, which will
also be smeared by the doppler and relativistic effects. One may
expect the absorption edge to be observable, and that, since the edge
will also be highly redshifted, some of the line flux may fall into
the edge. Here we compute in detail the expected spectrum from the
RB97 model. It is important to do this because there is no clear
evidence for absorption in the MCG--6-30-15 data, and there is the
possibility that the smeared absorption edge may affect the observed
line profile.

\section{Rest frame reflection spectra}
The two main parameters in the RB97 model, namely the height of the
point source above the disc $h$, and the overall radiative efficiency
of the accretion disc $\eta_{\rm x}$, have to be finely tuned in order
to produce a line profile that agrees qualitatively with the data. We
adopt parameters typical of those determined by RB97 in which the
X-ray continuum, a power law of photon index 2, is assumed to
originate at a height $h=3.5m$ along the rotation axis of the disc. A
small proportion of these photons will reach a distant observer, but
many will be bent towards to disc and blueshifted, causing an
enhancement in the illuminating intensity seen by the accreting
matter. We have calculated this numerically to determine the
ionization parameter as a function of radius. Reynolds \& Begelman
(1997) note that the ionization state is independent of the mass
accretion rate and accretion disc viscosity law since the velocity is
approximated by the local free-fall value and the continuum luminosity
is a fixed fraction $\eta_{\rm x}$ of the total accretion power
extractable from a Schwarzschild black hole. The density of the
infalling matter follows their expression (18). In Fig.~1 we plot the
ionization parameter and the Thomson depth through the infalling
matter as functions of radius. Although we have not explored the
entire parameter space, choosing different values for these parameters
should not qualitatively alter our results since, in order to fit the
data, there must be an annulus of emitting material significantly
within $6m$ and this must also have an associated absorption
edge. There is little freedom in the choice of parameters $h$ and
$\eta_{\rm x}$ if the RB97 model is to produce a line similar to that
seen in the data.

\begin{figure}
\centerline{\psfig{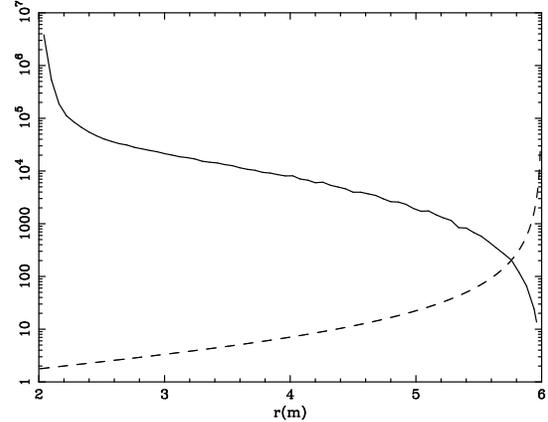}}
\caption{Ionization parameter $\xi$ (solid line) and optical depth
$\tau_e$ as functions of radius for a source height $h=3.5m$ and
efficiency $\eta_{\rm x}=0.003$.}
\end{figure}

The reflected X-ray continuum is computed in a frame in which the gas
is at rest. At a given radius the illuminating flux seen by the matter
in its rest frame and the ionization parameter are know, and we use
the method of Ross \& Fabian (1993) to compute the reflected
spectrum. This calculation takes into account the vertical ionization
structure, Compton scattering and resonant absorption within the
matter. The energy resolution is variable, with the greatest
resolution near the emission line features, and is significantly
higher than the broad line and edge features we wish to discuss
later. We have computed spectra for both solar and twice solar iron
abundances, and beyond $6m$ we have assumed the disc to be cold. The
spectra were calculated at 11 different radii, and we linearly
interpolate between them to obtain the spectrum at a particular
radius.  Fig.~2 shows the results of these calculations at a number of
radii for solar iron abundance. The ionization parameter increases
with decreasing radius and the four regions discussed above are
illustrated. Other spectral features can be seen below 6 keV, and are
due to the fluorescent lines of other metals.

\begin{figure}
\centerline{\psfig{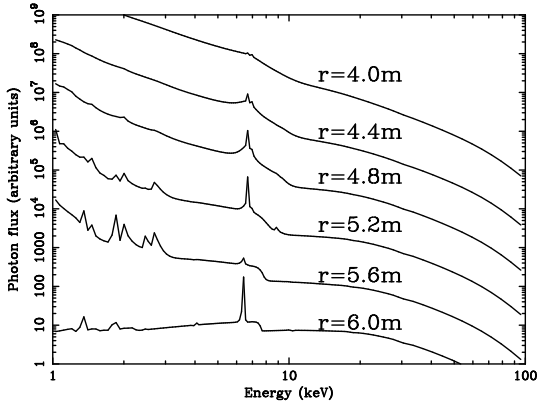}}
\caption{Reflected spectra
(offset for clarity), computed in a frame in which the gas is at rest,
for solar iron abundance. The iron line and absorption edge are
clearly seen to vary with radius, $r$.}
\end{figure}

\section{The observed spectrum}
To compute the predicted spectrum from the RB97 model we trace photon
paths backwards from a hypothetical observer through an array of equal
solid angle pixels until they hit the accretion disc or enter the
black hole. The observer is located at at a distance $r=1000m$ and at
an angle $\theta=30^\circ$ to the rotation axis of the disc, which is
thought to be the inclination angle of the disc in MCG--6-30-15. The
accretion disc is assumed to be geometrically thin and located in the
plane $\theta=\pi/2$, outside the black hole. The accretion flow is
divided into two distinct regions, that outside $6m$ where the
particles are assumed to follow Keplerian orbits, and that inside $6m$
where the particles are assumed to be freely falling into the black
hole from the radius of marginal stability. For photon paths that hit
the disc we can compute the spectrum that will reach the distant
observer from that point. To do this we make use of the fact that the
quantity $I(\nu)/\nu^3$ is invariant along the photon path, where
$I(\nu)$ is the specific intensity at a photon frequency $\nu$ (Misner
et al 1973, p588). We assume that the emission is isotropic, which is
a reasonable assumption in this case since the effect of limb
darkening is likely to be small as the disc is being viewed almost
face on, and the illuminating flux has a high angle of incidence onto
the disc (Cunningham 1976). If the point on the disc has an intensity
in the rest frame of the accreting matter of $I(\nu)$ at a frequency
$\nu$, and in propagating to the observer the photons are redshifted
such that a photon emitted at a frequency $\nu$ is observed at a
frequency $g\nu$, where $g=(1+z)^{-1}$ and $z$ is the redshift, then
the flux observed in a frequency interval $(\nu,\nu+\Delta\nu)$ in a
solid angle $\Delta\Omega$ is then given by

\begin{equation}
F(\nu,\nu+\Delta\nu)=\frac{\Delta\Omega}{4\pi}g^3I(\nu/g).
\end{equation}

This takes into account the transformation of the frequency element,
and if $I(\nu)$ is replaced by a delta function at a frequency $\nu_0$
in the rest frame of the matter we recover the usual $F\propto
g^4\delta(\nu-g\nu_0)$.

By integrating over the entire surface of the disc the overall
predicted spectrum from the RB97 model may be calculated. We have
computed the spectrum in the energy range 1--20 keV where we expect
little contribution from photon energies outside the 0.02--100 keV
range in which the rest-frame spectra have been calculated. The small
direct contribution from the illuminating source itself has also been
included.

The spectrum that we predict for the RB97 model is shown in Fig.~3,
which appears to consist of a power law with an extremely broad iron
line, an absorption edge, and a Compton reflection component. This
particular calculation has been performed for twice solar iron
abundance. The precise photon index of the power law component is
dependent upon the range of energies over which we fit the predicted
spectrum, but is found to be in the range $\sim$ 1.9--2.1. Fig.~3 also
shows the iron line profile computed using the same method as Reynolds
\& Begelman (1997) in which the iron line is treated as a delta function in
the rest-frame of the gas and the other features of the reflected
continuum are not considered. Their line profile has been added to a
power law for a simpler comparison, and no absorption or Compton
reflection components are present. It can also be seen from this
figure that the line profile that we predict for their model differs
from the profile that Reynolds \& Begelman (1997) have calculated.

\begin{figure}
\centerline{\psfig{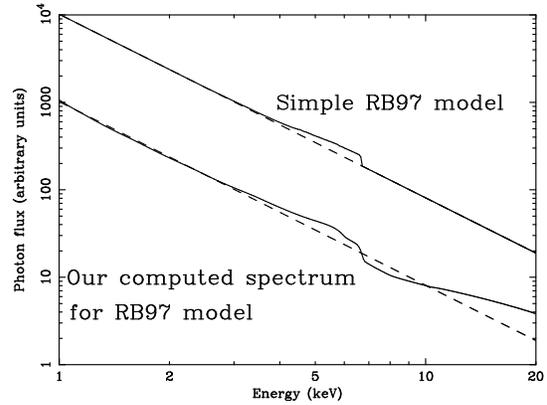}}
\caption{A comparison of the spectrum we expect from the RB97 model
in which we have computed the reflected continuum in detail in the
rest-frame of the matter (lower curve), with a spectrum obtained by
adding a power law to the line profile we expect if the reflected
continuum in the rest-frame of the matter is approximated by a delta
function representing the fluorescent line emission, and the other
features of the reflected continuum are ignored (upper curve).}
\end{figure}

To obtain a line profile that we can compare with that seen in the
data we have fitted our predicted spectrum with a power law between
2--10 keV to obtain the line profile of Fig.~4. For twice solar iron
abundance the best fitting power law has a photon index of
$\Gamma=1.94$. The iron line profile that we obtain is extremely
broad, highly redshifted, and appears to have a bite removed from its
high energy end. This bite may be attributed to some of the line flux
having fallen into the smeared absorption edge. If we consider the
spectrum from the simple RB97 model which approximates the iron line
component of the reflected rest-frame spectrum without any of the
other features then it does not predict an absorption edge or a bite
from the line. In Fig.~4 we have also overlayed the results for solar
iron abundance, the effect of which is to reduce the equivalent width
of the iron line and the optical depth of the absorption edge.

\begin{figure}
\centerline{\psfig{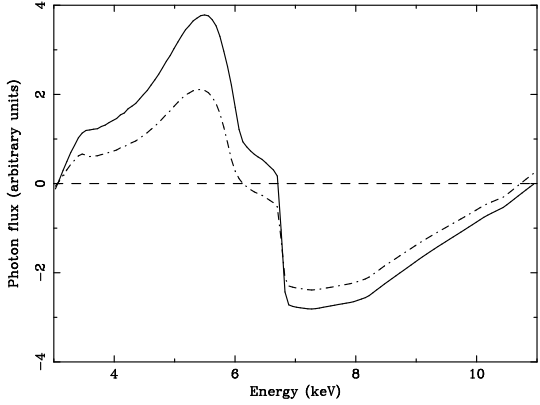}}
\caption{Our predicted line profiles for the RB97 model taking into
account the full reflected continuum, for both solar (dot-dashed) and
twice solar (solid) iron abundance. A best fitting power law has been
subtracted from our computed spectrum.}
\end{figure}

If we ignore the spectrum outside the range 3--7 keV then the
measurable quantities such as the equivalent width of the iron line
and the optical depth of the absorption edge are sensitively dependent
on the photon index of the power law that is fitted over this energy
range. Consider, for example, increasing the photon index of the power
law that we have fitted to our predicted spectrum of the RB97 model
shown in Fig.~3. This would result in the measured equivalent width of
the iron line increasing, and the measured optical depth of the
absorption edge decreasing. Since the energy range 3--7 keV contains
both the broad iron line and the absorption edge the true level of the
continuum at a given energy is difficult to estimate, and we would
have to consider the spectrum outside this energy range to be able to
rule out such a photon index with any certainty. If we now consider
Fig.~4, we can see that in fitting a power law to the predicted
spectrum between 2--10 keV it is possible for the photon index of the
power law to be overestimated, since the deficit of flux above $\sim
6.5$ keV due to the absorption edge would lead to an underestimate of
the true value of the continuum. We may face similar problems when
considering the actual X-ray data if we are forced to use a restricted
range of energies.

\section{MCG--6-30-15}

Observations of the Seyfert 1 galaxy MCG--6-30-15 during a deep
minimum of its light curve showed the iron line to broaden and extend
to lower energies than at other times (Iwasawa et al 1996). This
suggested that the source of the iron fluorescence moved inwards so
that the observed line had a greater redshift. The best-fitting model
for the 0.4--10 keV ASCA SIS spectrum in this interval, using two
ionized oxygen absorption edges (Otani et al 1996) and a Laor (1991)
line for a maximal Kerr black hole, has a power law with a photon
index of $\Gamma=1.75^{+0.08}_{-0.03}$. If we ignore the energy
regions with warm absorption and iron emission and just fit the
spectrum jointly in the 0.4--0.5, 2.5--3.0 and 7.0--10.0 keV bands
then $\Gamma=1.71\pm0.4$ (reduced $\chi^2=1.2$). A similar result is
also obtained if we just fit the spectrum in the 2.5--3.0 and
7.0--10.0 keV bands.

In Fig.~5a we have plotted our computed line profile for the RB97
model alongside the X-ray data for the iron line seen in MCG--6-30-15
during the deep minimum of its light curve. To produce our predicted
line profile a best fitting power law between 2--10 keV of photon
index 1.94 has been subtracted from our predicted spectrum with solar
iron abundance, and to produce the line profile seen in the data a
best fitting power law of photon index 1.75 has been subtracted from
the data. We see that the predicted line profile possesses a large
absorption edge beyond 6.9~keV, which is not present in the data.  We
therefore conclude that the absence of an iron edge in the data is
inconsitent with our calculated spectrum for the physical model of
RB97.

In Fig.~5b we show the effect of subtracting a power law with a
greater photon index of $\Gamma=2.05$ from our predicted spectrum to
produce a line profile. As we noted earlier this has the effect of
reducing the measured optical depth of the absorption edge. To produce
such a fit, however, we would have to assume that the photon index of
the power law source decreases even more significantly during the flux
minimum of the light curve. This would then cause our computed
spectrum to significantly underpredict the flux below the oxygen
absorption edges (i.e. below 0.7 keV), by approximately one third. If
soft excess emission were used to account for this deficit in flux
then it should be detectable in detailed photo-ionization model fits
to the warm absorber. Such soft excess emission is not required in
fits to the warm absorber in MCG--6-30-15.

\begin{figure}
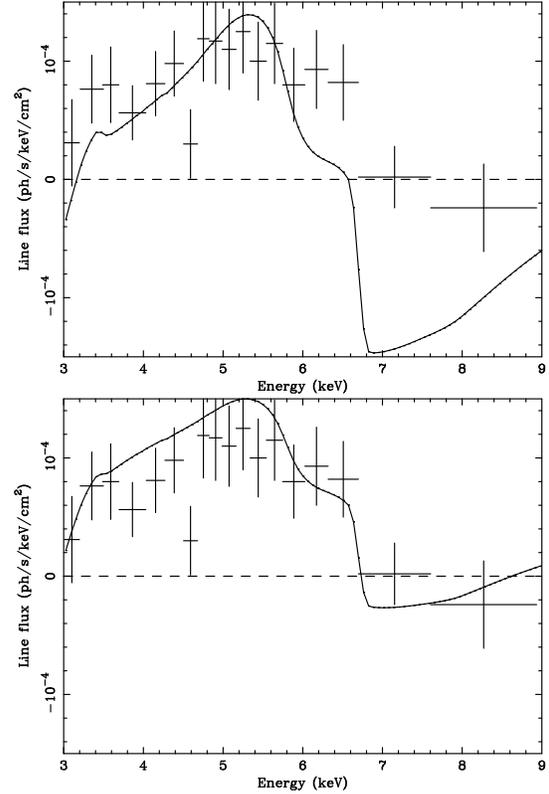

\centerline{\psfig{figure=new_fig5a.ps,width=0.4\textwidth,angle=270}}
\centerline{\psfig{figure=new_fig5b.ps,width=0.4\textwidth,angle=270}}
\caption{Comparisons of our predicted RB97 line profile with the
observed data for MCG--6-30-15. In Fig.~5a (upper panel) it is assumed
that the underlying power-law index has been accurately measured
whereas in Fig.~5b (lower panel) the observed line profile is compared
with the pseudo-line created from both the emission and absorption
features. This last solution requires an underlying power law have a
lower photon index than that observed.}
\end{figure}

\section{Discussion}
We have computed in detail the X-ray spectrum that we would expect to
see from the RB97 model for the extremely broad and highly redshifted
iron line seen during a deep minimum of the light curve of
MCG--6-30-15.  Our computation makes use of the full reflected X-ray
spectrum including the fluorescent iron line and absorption edge. We
predict that the overall observed spectrum should possess an
absorption edge, which is inconsistent with the data. We also obtain a
different line profile, although with present X-ray telescopes
detailed analysis of the shape of the line profile is not possible. It
is possible to provide an improved fit to the data but this requires
the power law of the X-ray source to have a photon index that is
significantly lower than 2. We conclude that the data favour a Kerr
over a Schwarzschild black hole model.

Nevertheless, the work of Reynolds \& Begelman (1997) has highlighted
the potential importance of flows within the marginally stable
orbit. If the irradiating X-ray source is at some distance above the
disc, particularly in a central location, then fluorescent iron line
features from such inflowing material should be observable. That the
predicted deep absorption edge is not seen in the case of MCG--6-30-15
supports a Kerr model and, since most of the time the line is not so
broad, argues that the irradiating source lies close to the disc and,
at times, changes in radius.

The edge in the predicted spectrum of the RB97 model is particularly
large because the material is ionized, so enhancing the contrast at
the edge. It may be possible to have less highly ionized material
within $6m$ if the accretion disc were gas pressure rather than
radiation pressure dominated all the way down to $6m$ since then the
density of the disc would be many times larger. There is also a small
edge to the cold disc reflection expected in the standard model. It is
unlikely that such a small edge would be detectable with current X-ray
telescopes, however.

It is also useful to consider other methods of probing the
innermost regions of the accretion disc. If we assume rapid large
variations in the continuum are due to flares over the disc, then
the time delay between the continuum variation and the fluorescent line
response from the disc may be used to estimate the height of the flare
above the disc. In the case of the RB97 model this would need to be
larger than that for a coronal model in order to provide sufficient
illumination of the material within $6m$. As well as the time delay
between the continuum change and the response of the disc, the
evolution of the iron line profile with time can tell us about the
geometry of both the source and the disc. Unfortunately with present
X-ray telescopes the photon flux in the iron line is only a few
hundred counts per day, and integration times are necessarily so large
that observations of short term variability are unfeasible.

It is exciting that we are now debating and able to distinguish gross
details of the accretion flow of matter at radii less than
$6m$. Future observations with ASCA, AXAF, XMM, ASTRO-E and
Constellation-X will continue this exploration of the very near
environment of black holes.

\section*{ACKNOWLEDGEMENTS}
AJY and ACF thank PPARC and the Royal Society for support,
respectively.

\end{document}